\documentclass[12pt,preprint]{aastex}

\slugcomment{}

\shorttitle{Photospheric Bright Points}
\shortauthors{P.H. Keys et al.}

\begin{document}

\title{The Velocity Distribution of Solar Photospheric Magnetic Bright Points}

\author{P. H. Keys, M. Mathioudakis, D. B. Jess, S. Shelyag, P. J. Crockett}
\affil{Astrophysics Research Centre, School of Mathematics and Physics, Queen's University, Belfast, BT7~1NN, 
Northern Ireland, U.K.}
\author{D. J. Christian}
\affil{Department of Physics and Astronomy, California State University, Northridge, CA 91330, USA}
\author{\and F. P. Keenan}
\affil{Astrophysics Research Centre, School of Mathematics and Physics, Queen's University, Belfast, BT7~1NN, 
Northern Ireland, U.K.}
\email{pkeys02@qub.ac.uk}

\begin{abstract}
We use high spatial resolution observations and numerical simulations to study the velocity 
distribution of solar photospheric magnetic bright points. The observations were obtained with the 
Rapid Oscillations in the Solar Atmosphere instrument at the Dunn Solar Telescope, while the 
numerical simulations were undertaken with the MURaM code for average magnetic fields of 
200~G and 400~G. We implemented an automated bright point detection and tracking algorithm 
on the dataset, and studied the subsequent velocity characteristics of over 6000 structures, 
finding an average velocity of approximately 1~km{\,}s$^{-1}$, with maximum values of
7~km{\,}s$^{-1}$. Furthermore, merging magnetic bright points were found to have 
considerably higher velocities, and significantly longer lifetimes, than isolated structures.
By implementing a new and novel technique, we were able to estimate the background 
magnetic flux of our observational data, which is consistent with a field strength of 400~G.
\end{abstract}


\section{Introduction}
\label{Intro}
Granulation forms the instantly recognisable ``patchwork'' pattern of the solar photosphere and 
dominates quiet Sun regions. The dark intergranular lanes are formed as a result of convective 
downflows and are the regions where magnetic bright points (MBPs) can be observed. It is the 
horizontal transfer of magnetic flux from the centre of granules, and into the lanes, that forms MBPs. 
In the MBPs, the magnetic flux clumps together to form small magnetic concentrations with field 
strengths of the order of a kiloGauss \citep{Sten85, Solan93}. These are some of the 
smallest features currently observable on the solar surface, and due to their dynamic nature, 
can provide a conduit for the transfer of kinetic energy into the upper solar atmosphere 
\citep{deWi09}. 

Due to the increased computational power available in recent years, significant progress has been made in 
understanding the complex interplay between convective energy transport and radiation in these 
concentrated magnetic elements. MBPs are best observed as strong intensity enhancements in 
G-band intensity images \citep{Mull84, Ber96} which have intensity 
peaks between $0.8-1.8$ times the mean photospheric intensity \citep{San04, Lan02}. 
These enhancements are due to an increase in continuum intensity, caused by the 
continuum formation layer being depressed into the deeper, hotter layers of the solar photosphere, 
where regions of strong magnetic field concentrations and reduced CH abundances exist 
\citep{Stein01, Shel04}.

The significance of MBPs on solar atmospheric energy transport and dynamics has led to several 
studies on their area coverage and size distribution.  
\citet{San04} found that MBPs have a minor axis of 135~km, while \citet{Wie04} determined a predominant 
diameter of $160 \pm 20$~km. In a recent study, \citet{Croc10} utilised high spatial resolution 
observations and magneto-hydrodynamic (MHD) simulations to conclude that the MBP area 
distribution peaks at $\approx$45000~km$^2$, with a sharp decrease in occurrence for smaller surface 
areas. The area of the smallest MBPs is defined by the width of the inter-granular lanes, which is in turn 
limited by the balance of radiative and convective energies in the magneto-convective processes.  

MBP velocities are induced by the expansion of granules, and are typically in the range 
of $1-3$~km{\,}s$^{-1}$ \citep{Ber96, Utz10}. It has been shown by \citet{Choud93} 
that magnetic footpoints with velocities greater than $\sim$2~km{\,}s$^{-1}$ can 
excite magneto-sonic kink waves, which can transport sufficient energy to heat the 
localised quiet corona, under a two layer atmospheric approximation. Also, recent observations 
\citep{Bon08, Wed09} and simulations \citep{Shel11} 
suggest the presence of vortex motions in the photosphere, with a possible connection to the spiral 
tracks of MBPs and chromospheric swirls. Very recent 
studies by \citet{shelyag3} and \citet{fedun11} have indicated how vortex motions 
generate a significant amount of Poynting flux directed outwards from the photosphere, 
and, as a result, may be the source of various observed MHD wave modes.


In this paper, we use high spatial and temporal resolution observations, in addition to 
numerical simulations, to determine the velocity distribution of a large sample of MBPs. 
The observations and numerical simulations are described in \S~\ref{obs}, while 
the methodology used, and the values obtained for the velocities of MBP structures, 
are detailed in \S~\ref{analy}. As our tracking algorithm can detect and monitor 
bright point chains, as well as isolated brightenings and merger events, we believe 
that this is a unique study of the dynamics of MBPs in the solar photosphere. 
Differences between the velocity characteristics of non-merging MBPs, and those that 
undergo mergers with other bright points, are discussed in \S~\ref{analy}. 
Finally, our concluding remarks are given in \S~\ref{conc}.

\section{Observations and Numerical Simulations}
\label{obs}
The data employed in this study were obtained using the Rapid Oscillations in the Solar Atmosphere 
\citep[ROSA;][]{Jess10} instrument, which is installed as a common-user facility at the 
76~cm Dunn Solar Telescope (DST), in New Mexico, USA. Observations were obtained during a 
period of excellent seeing on 2009 May 28, using a $9.2$~{\AA} wide filter 
centred at $4305$~{\AA} (G-band). We observed a $70'' \times 70''$ quiet Sun region at disk centre 
for $\sim$50~minutes, achieving diffraction-limited imaging with $0''.069$~pixel$^{-1}$.
The images were reconstructed using Speckle algorithms \citep{Wog08}, while image de-stretching 
was performed using a $40 \times 40$ grid \citep[equating to a $\approx1''.7$ separation between 
spatial samples;][]{Jess07, Jess08}. These processes were implemented to 
remove the effects of atmospheric seeing from the dataset. G-band images were taken at a raw 
cadence of 0.033~s, while after speckle reconstruction the cadence was reduced to 0.528~s. 
Reconstructed images were then binned into consecutive groups of four to improve the 
signal-to-noise and reduce the overall volume of the dataset, providing a final 
image cadence of 2.1~s. 

Simulated G-band images were produced using the detailed radiative transport technique 
described by \citet{Shel04}, with the solar photospheric magneto-convection models for the 
radiative transport calculations provided by the MURaM radiative MHD code \citep{Vog05}. A 
computational domain of size $12\times12\times1.4$~Mm$^3$, was employed for the simulations,
resolved by $480\times480\times100$ grid cells, providing a horizontal two-pixel resolution of 
50~km. The level corresponding to the visible solar surface is located approximately 600~km 
below the upper boundary of the domain. Side boundaries of the domain are periodic, while 
the upper boundary is closed for vertical and stress-free horizontal plasma motions and the 
bottom boundary is transparent. The numerical solution is stabilised against numerical 
instabilities using hyperdiffusive source terms \citep{CauKo01, Vog05, Shel08}.

In the initial stage of the numerical simulation, we use a well-developed non-magnetic photospheric
convection snapshot, where we introduce uniform unipolar vertical magnetic fields of 200~G and
400~G, representing two different levels of solar magnetic activity. Following the injection of 
magnetic fields into the simulation, they quickly become horizontally transfered into the intergranular lanes, 
creating magnetic field concentrations with a strength of up to 2~kG at the photospheric level. 
After a few convective turnover timescales, we record 219 and 177 magneto-convection 
snapshots, at 200~G and 400~G, respectively, for further analysis. To obtain the G-band intensities emerging 
from the simulated photospheric snapshots, a spectral band in the wavelength range from 4295~{\AA} to 
4315~{\AA}, which contains 241 CH molecular lines and 87 atomic lines, was calculated for each vertical ray 
in the snapshots, and convolved with a filter function centred at 4305~{\AA} with a width of 9.2~{\AA}.

\section{Analysis and Results}
\label{analy}
The MBPs analysed in this paper were detected using the algorithm developed by \citet{Croc10}. 
This automated algorithm uses the intergranular lanes to produce a binary map of bright features. 
A compass search is employed to disentangle features within the lanes, i.e. MBPs,  from bright granules. 
Finally, intensity thresholding across the feature is used to determine more accurately the size and 
shape of the MBP. Long-lived features are then established, and stabilised, by examining subsequent 
frames to observe the MBP's evolution in time. More details of these procedures may be found in \citet{Croc10}. 

\begin{figure*}[h!]
\centering
\includegraphics[width=16.5cm]{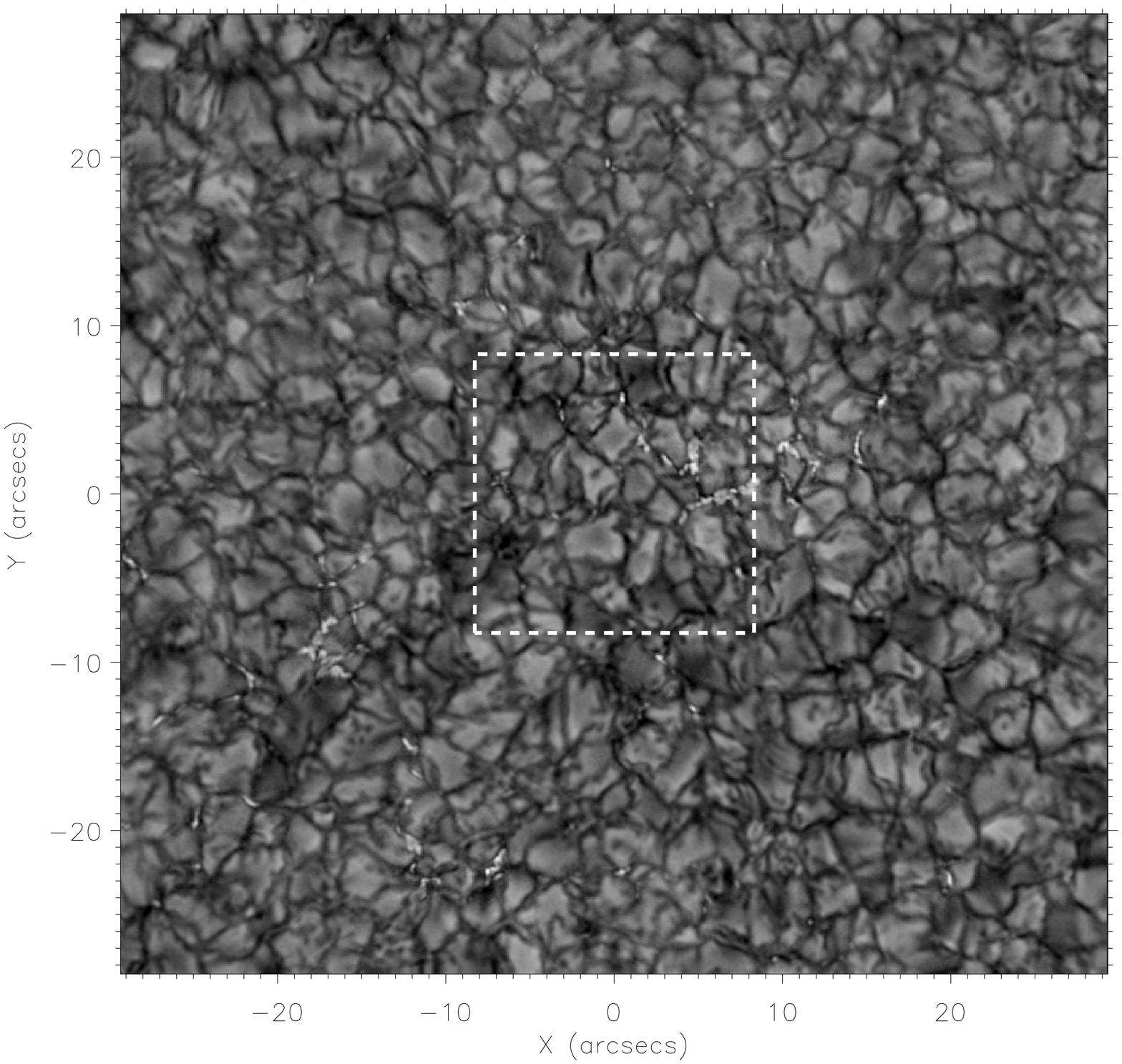}
\caption{The full $70''\times70''$ ROSA field-of-view of the observational G-band data used in this study. 
The dashed white box indicates a $16''\times16''$ region which is expanded in Figure~\ref{Fig2}(a). 
The axes are in heliocentric arcseconds, with (0,0) representing the centre of the solar disk.}
\label{Fig1}
\end{figure*}

\begin{figure*}[h!]
\centering
\includegraphics[width=16.5cm]{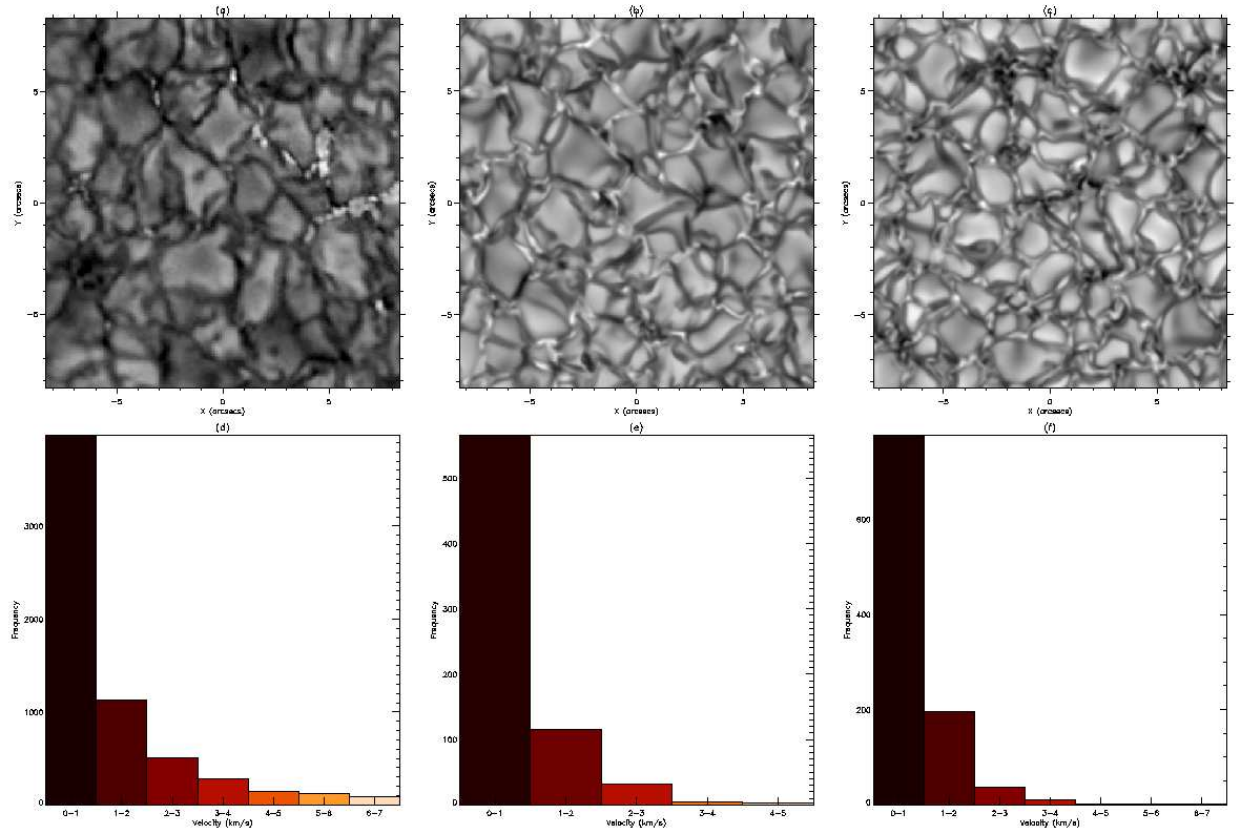}
\caption{The upper panels contain $16''\times16''$ snapshots of the three datasets used in this study. 
(a) is an expanded view of the boxed region in Figure~\ref{Fig1}, while (b) and (c) represent the 200~G 
and 400~G simulations, respectively. The velocity distributions for each dataset 
are shown below their corresponding snapshot in panels (d--f).}
\label{Fig2}
\end{figure*}

Figure~\ref{Fig1} shows the full ROSA field-of-view, while Figure~\ref{Fig2}(a) shows an expanded 
version of the $16''\times16''$ region highlighted in Figure~\ref{Fig1}, together with sample 
images of the 200~G (Fig.~\ref{Fig2}b) and 400~G (Fig.~\ref{Fig2}c) simulations. 
The algorithm produces an outline of the centre of gravity of each MBP observed, as well 
as a direct measurement of the area of each feature. A 30~s cut-off was employed to remove 
those MBPs which are short-lived and unlikely to have a significant horizontal velocity component, 
as well as assisting the removal of any false detections that may adversely affect the results. 
Figure~\ref{Fig3} shows snapshots from two videos that have been included as an online resource. 
The first depicts the tracking capabilities of the algorithm for an individual MBP throughout 
its 336~s lifetime, while the second displays how an MBP from the 200~G MHD simulation is tracked.

Our observed G-band dataset included a total of 6236~MBPs with lifetimes exceeding 30~s. 
These MBPs were found to have an average velocity of approximately 1~km{\,}s$^{-1}$, 
and a mean lifetime of 91~s. The highest velocity recorded in this study was 7~km{\,}s$^{-1}$, 
a value comparable to the photospheric sound speed, while the longest MBP lifetime observed 
was approximately 20~minutes. Similar values were obtained when the same methodology 
was applied to the numerical simulations. A total of 721 MBPs were detected in the $16''\times16''$ 
200~G simulation, with an average velocity of 0.6~km{\,}s$^{-1}$, a maximum of 
4.1~km{\,}s$^{-1}$, an average lifetime of 108.2~s, and a maximum of just over 11~minutes. 
In the 400~G simulation, we detected 1024~MBPs, with an average velocity of 0.65~km{\,}s$^{-1}$ 
and a maximum of 6.2~km{\,}s$^{-1}$, an average lifetime of 168.4~s and a maximum 
of nearly 17~minutes. Histograms indicating the distributions of velocities for all three 
data sets are shown in Figure~\ref{Fig2}(d--f), with the results summarised in Table~1. 
A dependence of the velocity distribution on the average magnetic field strength 
(and thus on the solar magnetic activity level) suggests an alternative, and novel, way of 
estimating the net magnetic flux in the solar photosphere. Hence, as the best 
agreement with our observations is found for the 400~G simulation, we can infer the average 
magnetic flux of our observations as being close to 400~G. 
Small discrepancies between our observed and simulated datasets may be due to the large 
MBP sample present in our observations (6236 versus 1024). 

The distribution of velocities, for both the observed and simulated datasets, are shown as histograms in 
Figure~\ref{Fig2}(d--f). These demonstrate that the majority of MBPs have velocities between 0 
and 1~km{\,}s$^{-1}$, while some MBPs (i.e. 30\% in the observations, 6\% in the 200~G 
simulated data and 6\% in the 400~G) have velocities in excess of 2~km{\,}s$^{-1}$. 
The considerable fraction of MBPs with velocities greater than 2~km{\,}s$^{-1}$ is important, 
as rapid bursts in their motion may induce magnetosonic kink waves \citep{Choud93}. As shown previously, 
these kink waves may act as a conduit for imparting energy into the upper solar atmosphere. 

Although there is considerable agreement between the results obtained in the observed and 
simulated data, there are some slight discrepancies that need to be addressed for a more complete 
picture. For example, the maximum velocity, the maximum lifetime and the proportion of MBPs with 
velocities greater than 2~km{\,}s$^{-1}$ are all smaller in the simulated datasets. A possible 
reason is that the net (or average) magnetic field applied within the computational 
domain remains constant throughout the 200~G or 400~G time series, whereas the flux present 
in the observed data may constantly change with time. This dynamic nature of flux generation 
and annihilation could directly lead to the higher velocity distributions detected for the observed MBPs.

An alternative algorithm based on local correlation tracking \citep[LCT;][]{Wel04, Mat10} was 
applied to the observations and simulated time series, and yielded similar velocity characteristics to 
within 0.3~km{\,}s$^{-1}$.  
A significant number of MBPs, $\approx$21\% of those detected in the observations, were involved 
with a merging event. Only a few of mergers (31) are observed in the 200~G simulation, 
whereas 207 are detected in the 400~G simulation. Comparisons with an active region area 
would be a useful study in the future to observe how the frequency of merging events in such a 
feature compares to the quiet Sun region discussed here. 

There is only a slight difference between the velocities obtained for merging events and those of 
isolated MBPs. The average velocities evaluated for all the merging events in the observations 
was found to be 1.6~km{\,}s$^{-1}$, slightly above the measured value for isolated MBPs. However, 
the average lifetime of the merging MBPs is 201.2~s, more than twice 
the value for non-merging MBPs. These results are echoed in the values 
for the 200~G simulated dataset, although only 31 mergers were witnessed out of the 721 
detections. Here, the average velocity of the merging MBPs is 0.9~km{\,}s$^{-1}$, 
with an average lifetime of 227~s. For the 400~G simulation time series, the average velocity 
was found to be 0.8~km{\,}s$^{-1}$, with an average lifetime of a merging MBP of 329~s 
(determined over a total of 207 merging events). These results show that there is little difference in 
the velocity of a merging MBP when compared with its non-merging counterpart. However, the former 
do appear to have longer lifetimes. This effect may arise due to an increase in stability of the 
MBPs when they undergo a merger. One possibility to account for the marginal elevation of velocity 
characteristics for merging MBPs is the barycenter method used to establish the 
merging velocity. During a merger, the shift of the centre of gravity as the MBPs unite may 
lead to a larger recorded velocity value. This may also explain why we see such a high 
maximum velocity in the observed data set. It is most likely an artefact of MBP geometry changes, 
which induce a considerable motion on the barycenter of the MBP structure. Thus, some velocities 
may be overestimated by this method. However, employing LCT methods on the same MBP 
groups indicate that a fraction of these structures do in fact demonstrate velocities 
comparable to the sound speed.

\section{Concluding Remarks}
\label{conc}
We have used high resolution observations, and state-of-the-art MHD simulations, to determine 
the velocity distribution of magnetic bright points (MBPs) on the solar surface. The average velocity 
established from our observations is 1~km{\,}s$^{-1}$, averaged over a total of 6236 MBP 
structures, with a typical lifetime of 91~s. A total of 721 MBPs were tracked in the 200~G 
numerical simulation, with an average velocity of 0.6~km{\,}s$^{-1}$ and a mean lifetime of 
108.2~s. The slight differences between the observed and 200~G simulated datasets can be 
attributed to the initial value of magnetic field applied to the MHD simulation. Using an initial 
magnetic field of 400~G, the subsequent numerical simulation yielded values closer to the 
observational distribution. The average velocity and average lifetime of the 1024 MBPs detected 
in the 400~G data is 0.65~km{\,}s$^{-1}$ and 168.4~s, respectively. When 
local correlation tracking (LCT) was applied to each data set, the difference in MBP velocities 
was found to be $\approx$0.3~km{\,}s$^{-1}$, showing a level of consistency between the 
two independent techniques. 

For the merging events witnessed in the observational dataset, the average velocity was found to be 
1.6~km{\,}s$^{-1}$, with a mean lifetime of 201.2~s. The 200~G simulated dataset only produced 31 
merger events, with an average velocity of 0.9~km{\,}s$^{-1}$ and mean lifetime of 
227~s. By contrast, the 400~G simulation had 207 mergers, with an average velocity of 0.8~km{\,}s$^{-1}$ 
and mean lifetime of 329~s. All of the data sets show that merging events exhibit longer average 
lifetimes, which we believe is a result of greater MBP stability as a result of the merger. Although 
the 200~G simulation produce similar MBP lifetimes, our data are most consistent with the 
400~G simulation in both appearance and in MBP merger lifetimes. We intend to expand on 
this study by observing the velocity characteristics of MBPs and merger events in the 
vicinity of an active region.

Importantly, we have presented a novel technique for estimating the average photospheric 
magnetic flux through comparison of observations and MHD simulations. By scaling the initial 
magnetic flux injected into the numerical simulation to replicate the MBP dynamics established 
in our observational time series, we can directly estimate the average background magnetic 
flux in the solar feature studied, without the necessity of vector magnetograms. Following this 
methodology, we find our observational data is consistent with an average magnetic flux of 
$\sim$400~G.

\acknowledgments
This work has been supported by the UK Science and Technology Facilities Council (STFC). 
Observations were obtained at the National Solar Observatory, operated by the Association of 
Universities for Research in Astronomy, Inc. (AURA), under cooperative agreement with the 
National Science Foundation. 
PHK thanks the Northern Ireland Department for Employment and Learning for a PhD studentship.
DBJ thanks the STFC for the award of a Post-Doctoral Fellowship.  
Effort sponsored by the Air Force Office of
Scientific Research, Air Force Material Command, USAF under grant number FA8655-09-13085

\clearpage

\clearpage

\clearpage
\begin{table}[]
\caption{Summary of MBP characteristics.}           
\label{table1}      
\centering                                    
\begin{tabular}{l c c c}         
\hline\hline                        
\textbf{Data Set} &  \textbf{ROSA Observations} &  \textbf{200G Sims} & \textbf{400G Sims}\\
\hline                                   
FOV size & $70''\times70''$ & $16''\times16''$ & $16''\times16''$\\
Average cadence (s) & 2.112 & 8.7 & 6.7\\
Number of MBPs detected & 6236 & 721 & 1024\\
Mean lifetime (s) & 91 & 108 & 168\\
Average velocity (km{\,}s$^{-1}$) & 1.0 & 0.6 & 0.65 \\
Maximum velocity (km{\,}s$^{-1}$) & 7.0 & 4.1 & 6.2\\
Percentage of detections that merge & 21\% & 4\% & 20\%\\
Average velocity of mergers (km{\,}s$^{-1}$) & 1.6 & 0.9 & 0.8 \\
Average lifetime of mergers (s) & 201 & 227 & 329\\
\hline                                            
\end{tabular}
\end{table}

\end{document}